\documentclass[11pt,twoside]{article}

\usepackage{asp2006}
\usepackage{epsf}
\usepackage{psfig}
\usepackage{lscape}
\usepackage{graphicx}
\begin{document}
\title{Gravitational heating, clumps, overheating}   %%% Fill in title
\author{Yuval Birnboim}   %%% Fill in author names
\affil{Harvard Smithsonian Center for Astrophysics. 60 Garden Street,
  Cambridge, MA 02138, USA}    %%% Fill in author affiliations

\begin{abstract} %%% Abstract to run on from here.
There is no shortage of energy around to solve the overcooling problem of
cooling flow clusters. AGNs, as well as gravitational energy are both
energetic enough to balance the cooling of cores of clusters. The
challenge is to couple this energy to the baryons efficiently enough,
and to distribute the energy in a manner that will not contradict
observational constraints of metalicity and entropy
profiles. Here we propose that if a small fraction of the baryons that
are accreted to the cluster halo are in the form of cold clumps, they
would interact with the hot gas component via hydrodynamic drag. We
show that such clumps carry enough energy, penetrate to the center,
and heat the core significantly. We then study the dynamic response of
the cluster to this kind of heating using a 1D hydrodynamic simulation
with sub-grid clump heating, and produce reasonable entropy profile in
a dynamic self-consistent way.
\end{abstract}

\section{Introduction}   
Galaxy clusters grow by accreting dark matter and baryons from their
surroundings. This is partly a continuous process of relatively smooth
accretion, and partly via mergers. The existence of a visible,
extended X-ray halo as well as limits on the size of the CD galaxy are
two indicators that the smooth, continuous accretion is the more
pronounced of the two. The baryons that are accreted in this process
settle in hydrostatic equilibrium within the cluster's potential
well. During this process they achieve virial equilibrium  by
passing through the virial shock and
converting the kinetic energy of the infall to thermal energy.
The baryons, at a temperature of a
few $keVs$, cool primary by emitting Bremsstrahlung radiation which
easily escape the halo and is observed by X-ray telescopes. 
After de-projection (tomography) of the luminosity and spectrum,
radial profiles of temperature and densities are derived, and one can
deduce the cooling times of the gas, which, for the centers of cooling
flow cluster, is $\le 1Gyr$, much shorter than the age of typical clusters
that, according to $\Lambda CDM$ formation history \citep{press74} should
have been in place at $z \ge 1$. Had this gas cooled, \emph{i)} cool gas
would have been seen in the halo (no gas below $T=\frac{1}{3}T_{vir}$
is observed), \emph{ii)} a census of all the baryons in galaxies is significantly
smaller than amount of gas that was expected to cool from the halo,
and \emph{iii)} the star formation within the CD would have
been $10^2-10^3M_\odot/yr$ (two orders of magnitude larger than
typically observed). These three contradictions are three manifestations of the overcooling
problem of cooling flow clusters. It is highly unlikely that there is
a ``hidden'' baryonic component in cluster halos, so most explanations invoke some kind of heating
mechanisms that would balance the cooling, keeping the gas hot
and diffuse.

The energies needed to compensate for the rapid cooling of cluster
halos are $\sim 10^{45}erg/sec$ which, over the lifetime of the
cluster amounts to $\sim 10^{62} erg$. These required energy rates can
originate from AGN emission  from the CD \citep[][as well as many others]{ciotti97}, and by gravitational
energy. Diffuse baryons falling into a gravitational well convert
gravitational energy to kinetic, and ultimately to thermal. This
thermal input into the system is usually local, and acts to heat the
infalling gas itself. It is necessary to couple this freshly accreted, hot
gas, with the central cooling gas. \citet{narayan01} have studied conduction and turbulence that could potentially couple
the external part to the inner halo. \citet{kim03}, as well as others,
deduce that although the amount of energy is sufficient, the
conduction coefficient is not enough and turbulence will produce the
wrong entropy and metalicity profiles. 

In this proceedings, we shall present a novel mechanism of
gravitational heating of
the central halo gas by the hydrodynamic interactions between cold
clumps of accreted gas and the halo gas \citep{db08} and
\citetext{Birnboim \& Dekel 2009, in preparation}. The structure of this
proceeding paper is as follows: First (section \ref{sec:energetics}) we show that the
accreting gas carries a sufficient rate of energy to compensate for
the cooling. Then we model infalling cold clumps, and study the physical processes of their interaction with
the diffuse gas, instabilities and survivability (section \ref{sec:physics}). Using this model, we study the
valid parameter space of these clumps (section \ref{sec:monte}). We further use these insights
to construct a sub-grid model for 1D hydrodynamic simulation and
present simulations of a cluster with and without such clumps (section
\ref{sec:hydro}). Finally
we discuss possible origins of these clumps, and summarize (section \ref{sec:summary}).

\section{Energetics}
\label{sec:energetics}
\begin{figure}[!ht]
\begin{center}
\scalebox{0.27}{\includegraphics*[.6in, 2.5in][8.in,10.in]{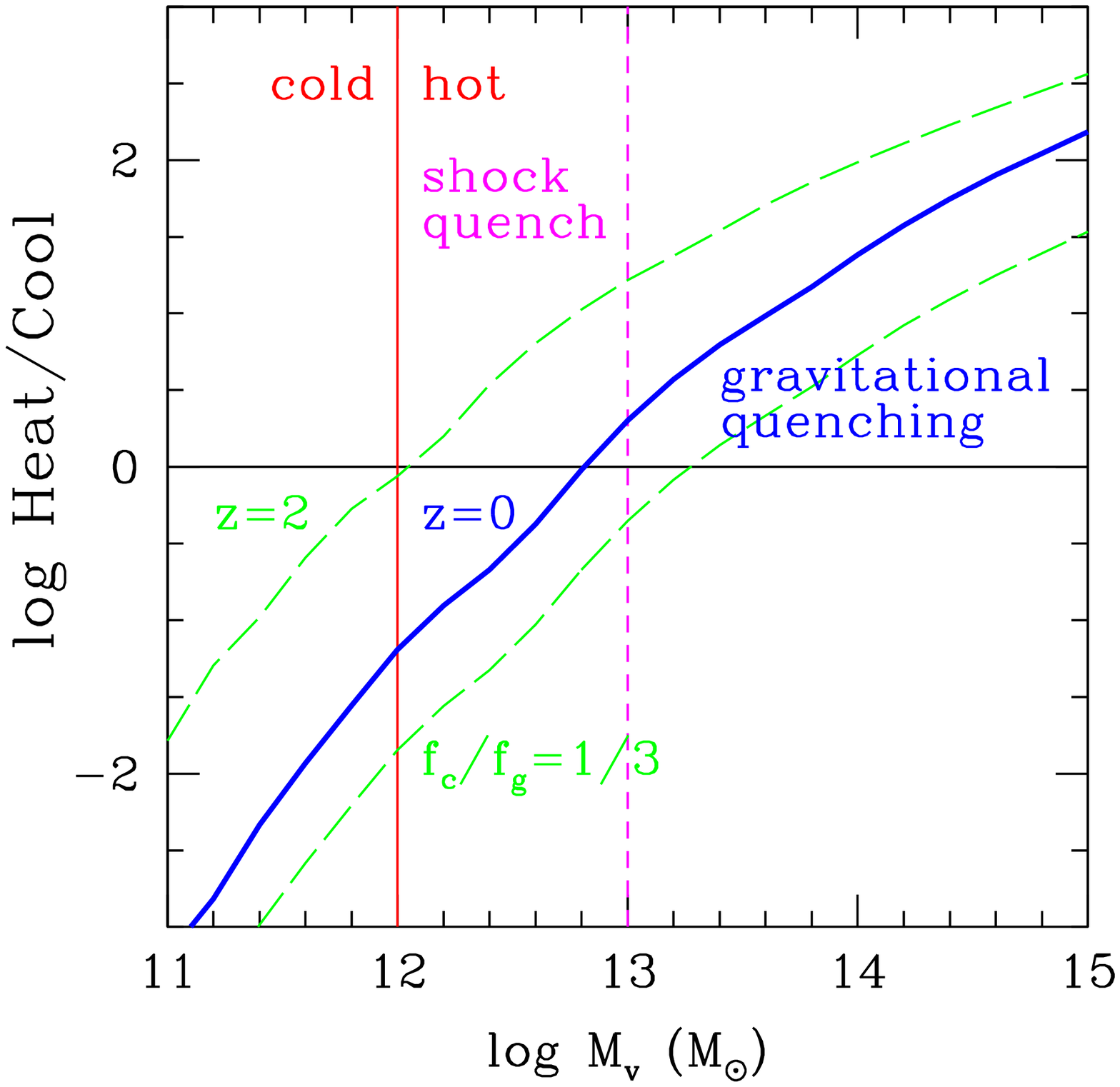}}
\scalebox{0.43}{\includegraphics*[0in, 6.5in][5.in,11.in]{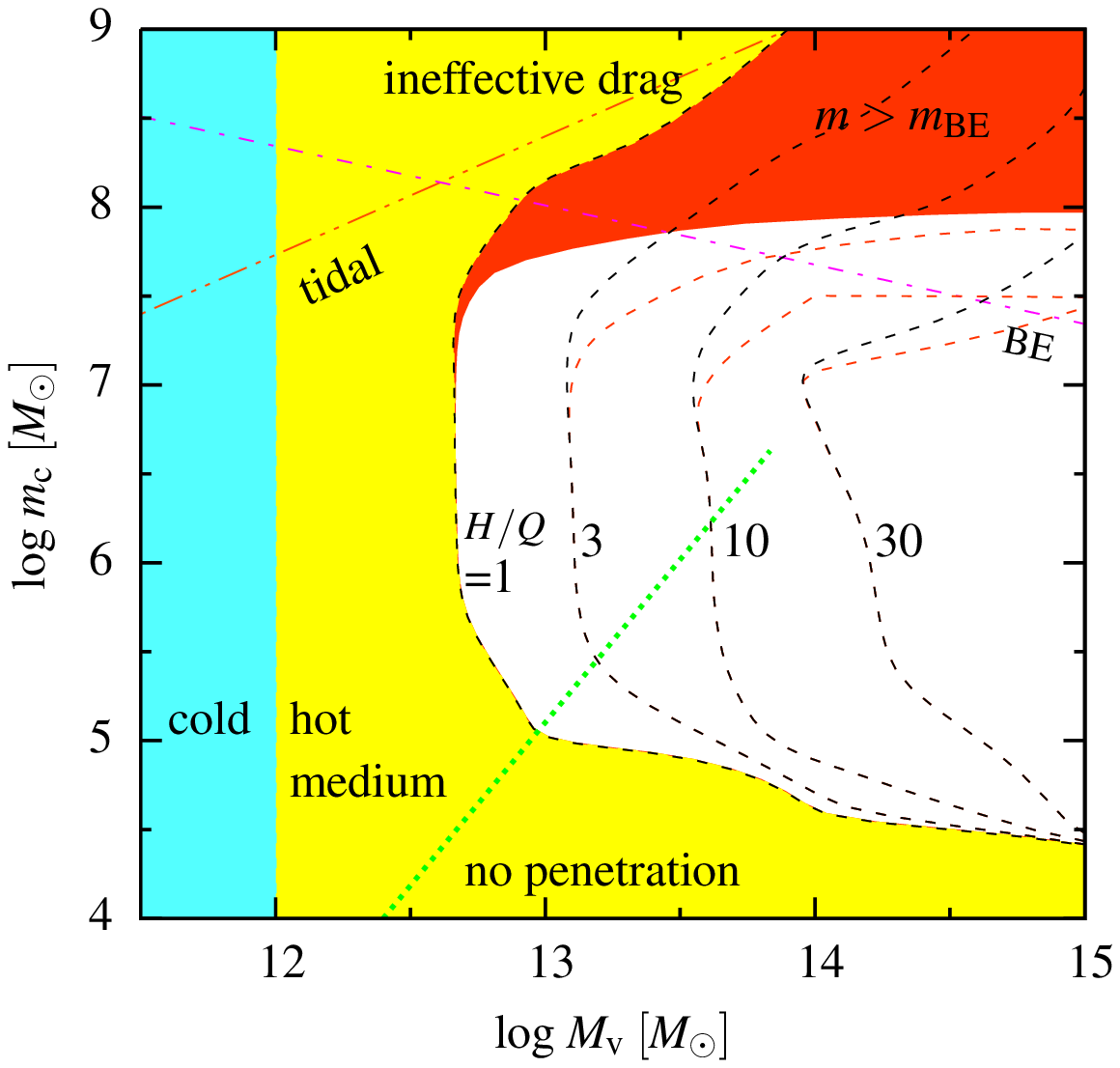}}
\end{center}
\caption{\emph{Left:} Heating/Cooling rate as a function of virial mass for the
  fiducial case (blue, solid line): $f_b=0.05$, $f_c=0.05$,
  $Z=0.3Z_\odot$ and $R_{final}=0.1R_{vir}$ at $z=0$. Similar lines for
  $z=2$ (upper,dashed green) and for $f_b=0.075$, $f_c=0.025$ (lower,
dashed green) are also presented. Below $M_{vir}=10^{12}M_\odot$ the
halo gas cannot be hot \citep{bd03,db06} and no hydrodynamic drag is
possible. \emph{Right:}
  allowed parameter space for clump heating mechanism on $m_c/M_{v}$
  space at $z=0$, $Z=0.3Z_\odot$ and $f_c=f_b=0.05$ : in white areas
  Heating/Cooling (H/Q) $>1$. The black lines show H/Q of $3,10,30$. The
  yellow areas have H/Q<1 and red areas are susceptible to
  Bonnor-Ebert instability. Both figures are originally from \citet{db08} by the kind
permission of John Wiley \& Sons Ltd.}
\label{fig:ener}
\end{figure}

Gas falling from infinity to the virial radius will carry specific
energy that correspond to the amount needed for heating itself to the virial
temperature. Here we assume that because the gas is clumpy, it can
penetrate further inwards than the virial radius, deeper into the potential well and
release more energy. For an NFW \citep{nfw97} halo profile with
typical cluster concentration \citep{bullock01_c}, gas penetrating
to $0.1R_{vir}$  will release $\sim 3.5$ times
the energy it would at $R_{vir}$ \citep{db08}. The baryonic profile is
a generalized NFW model \citet[see eq. 4 in][]{db08} with a core
($\alpha=0$) that is perhaps indicated by observations
\citep{donahue06} and hydrodynamic simulations
\citep{faltenbacher07,kaufmann08} \footnote{In later sections we will
  use a hydrodynamic simulation for the cluster formation and the need
  to assume a profile will become moot}. Figure \ref{fig:ener}, left, shows a
comparison between the global cooling rate \citep[using][and
$Z=0.3Z_\odot$]{sutherland93} and the heating rate assuming accretion
rate by \cite{neistein06,bdn07}, diffuse baryonic fraction $f_b=0.05$ and clump
fraction of $f_c=0.05$. The Heating/Cooling ratio becomes positive for
$M_{vir}\ge 10^{13}M_\odot$ so the gas accreted as clumps to clusters
is sufficient to counter the cooling. For clusters of $10^{15}M_\odot$
there is an overabundance of energy of two orders of magnitude, much
larger than typical energy emission rates from AGNs.

\section{Physics of cold clumps}
\label{sec:physics}
%drag
\emph{Drag forces.} The clumps in focus here are cold ($10^4K$) gaseous clumps, in
hydrostatic equilibrium with their surrounding hot gas. The
hydrodynamic drag force is:
\begin{equation}
f_{drag} =-\frac{\pi}{2}C_d \rho_g V_r^2 r_c^2 \hat{V_r},
\label{eq:drag1}
\end{equation}
with $\rho_g$ the ambient gas density, $V_r$ the relative velocity
between the clump and the ambient gas, $\hat{V_r}$ is the radius
vector of that velocity, and $r_c$ is the radius of the clump. $C_d$
is the drag coefficient, and for gaseous spheres it is of order of $1$ for
subsonic velocities and supersonic velocities alike \footnote{While
  the equation of drag is the same for the subsonic and supersonic case,
  the dissipation mechanism is different. For subsonic motions the
  sphere creates turbulence which dissipates to heat. For supersonic
  motions a bow shock travels in front of the sphere heating
  the ambient gas. In the transonic regime $C_d$ can increase to a
  factor of a few, but then decrease back to $C_d\sim 1$ for
  supersonic and subsonic velocities}. The drag force always act to
decrease the kinetic energy of the clump. This energy is converted
into thermal energy of the \emph{hot gas}  (see \citet{murray04} and
discussion in \citet{db08}). The rate of heating is $\frac{dE}{dt}=f_d
|V_r|$, and the trajectory of the clumps tends to become radial
because the velocity in the radial direction is replenished by the
gravitational force, but the tangential velocity decreases
monotonically. From eq. \ref{eq:drag1} it is evident that the
deceleration is more efficient for smaller clumps, larger velocities,
and larger gas density.

%survivability
\emph{Survivability of clumps.} Kelvin Helmholtz (KH) instabilities cause the clump to disintegrate after it repels its own mass in
ambient gas \citep{murray04}. The clumps disintegrate into a few pieces
(typically $2$), each piece undergoing the same drag forces and KH
instability. If the clumps are too small, conduction
will cause the clump to expand and disintegrate. The conduction
coefficient $f_s$ \citep{spitzer62}, depends strongly on the unknown magnetic
properties of the gas and is assumed here to be $0.01\le f_s\le
0.1$. A discussion on that is present in \citet{maller04}. If,
however, the clump is too massive, it will become gravitationally
unstable and collapse under its own gravity and the external
pressure. The Bonnor-Ebbert mass \citep{ebert55,bonnor56} sets the maximal
allowed clump mass.

%death of clumps.
\emph{Death of clumps.} As clouds fragment and become more susceptible to conduction and small
scale turbulence, heat from the surrounding hot gas will flow into the
cold clump residue, heating it until it joins its surrounding halo. In
this final stages, the clump ``steals'' energy from the hot
gas, cooling it. The net effect, however, is heating because the
clumps spends $3.5$ times than it ultimately absorbs. This cooling has been
taken into effect in the static and dynamic tests in the following
sections. In cases where clumps exceed the Bonnor-Ebert mass (Jeans
mass with external pressure confinement) the clump gas would turn into
star without such ``theft''.

%cooling
\emph{Local and global instabilities.} The cooling rate of the baryons, in the
Bremsstrahlung regime scales (for isobaric gas) scales as
$\frac{de_{cool}}{dt}\sim \rho_g^2T^{1/2} \sim \rho^{1.5}$. \citet{field65} noted that heating must
be scale with a power at least as large for the gas to be
cooling-stable.
However, the clump heating scales like $\rho_g$
(eq. \ref{eq:drag1}). This means that, assuming some parcel of gas
that is initially in heating/cooling equilibrium, if gas is perturbed
such that the cooling
becomes slightly faster than the heating, the gas contracts, causing
the discrepancy between
heating and cooling to increase even further. If, on the other hand,
heating is larger than the cooling, the gas will expand, causing the
cooling to become even less efficient with respect to the heating. As
we will show in section \ref{sec:hydro}, the later case actually
occurs, and causes some of the shells to heat and expand more than their
surrounding. This creates entropy inversions, which are regulated by
convection. This local instability, as it turns out, does
not effect the global stability of the halo: the location of the
virial shock, the temperature, and the dynamics of the gas throughout
the halo does not change as a result of this heating (see section \ref{sec:hydro}). 

\emph{Origin of clumps.} In section \ref{sec:monte} we show that
allowed masses for the gaseous clumps is between $10^5-10^8
M_\odot$. Mechanisms for formation of such clumps are still
under investigation. The virial temperature that corresponds to mini-halos with
baryonic mass $\le 10^8M_\odot$ is smaller than $10^4K$ so even if
dark matter mini-halos are formed, they cannot retain their gas after
reionization. Cooling instability, especially in filaments or partly
shocked gas at the edges of the clusters ($T\sim 10^5-10^6K$) can
potentially create such clumps. These clumps can also be produced
within halos \citep{maller04}. No hydrodynamic simulation today has
the resolution to investigate the formation of such clumps.

\section{Numerical tests and simulations}
\subsection{Static Monte-Carlo simulations}
\label{sec:monte}
The drag force equation and clump fragmentation have been incorporated
into a Monte-Carlo simulation of a static NFW halo with a baryonic
core \citep[see][for details]{db08}. In these simulations the
initial trajectories of 4000 clumps were drawn from reasonable distributions, and the
heating rates were calculated as a function of radius, following
fragmentation, evaporation, conduction, local Bonnor-Ebert masses, as
well as dynamical friction. In some cases, a dark matter (DM) counterpart
for the clumps was initially assumed. The DM subhalo generally gets
ripped from the baryonic part during the first passage near the
center, because the drag forces act on the baryons alone, and
introduce forces between the DM and the baryons.
Figure \ref{fig:ener}, right, describes results of the Monte-Carlo
approach by mapping the allowed range
in $m_c/M_v$ parameter space. \emph{For halos above $10^{13}M_\odot$, and
clumps between $10^5-10^8M_\odot$, the heating rate is larger than the
cooling rate}, for the fiducial values described in the caption.

\subsection{1D hydrodynamic simulations}
\label{sec:hydro}
\begin{figure}[!ht]
\plottwo{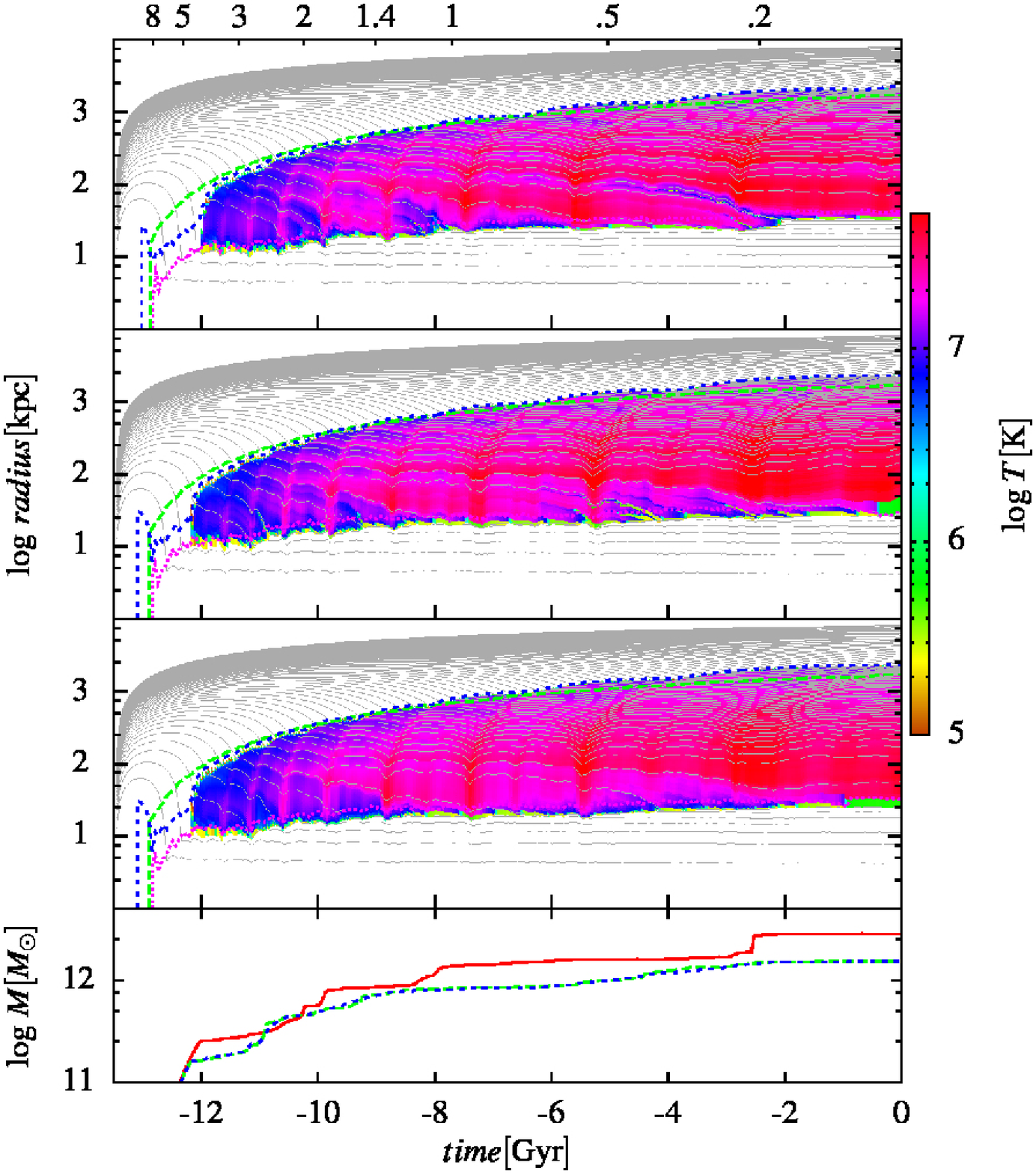}{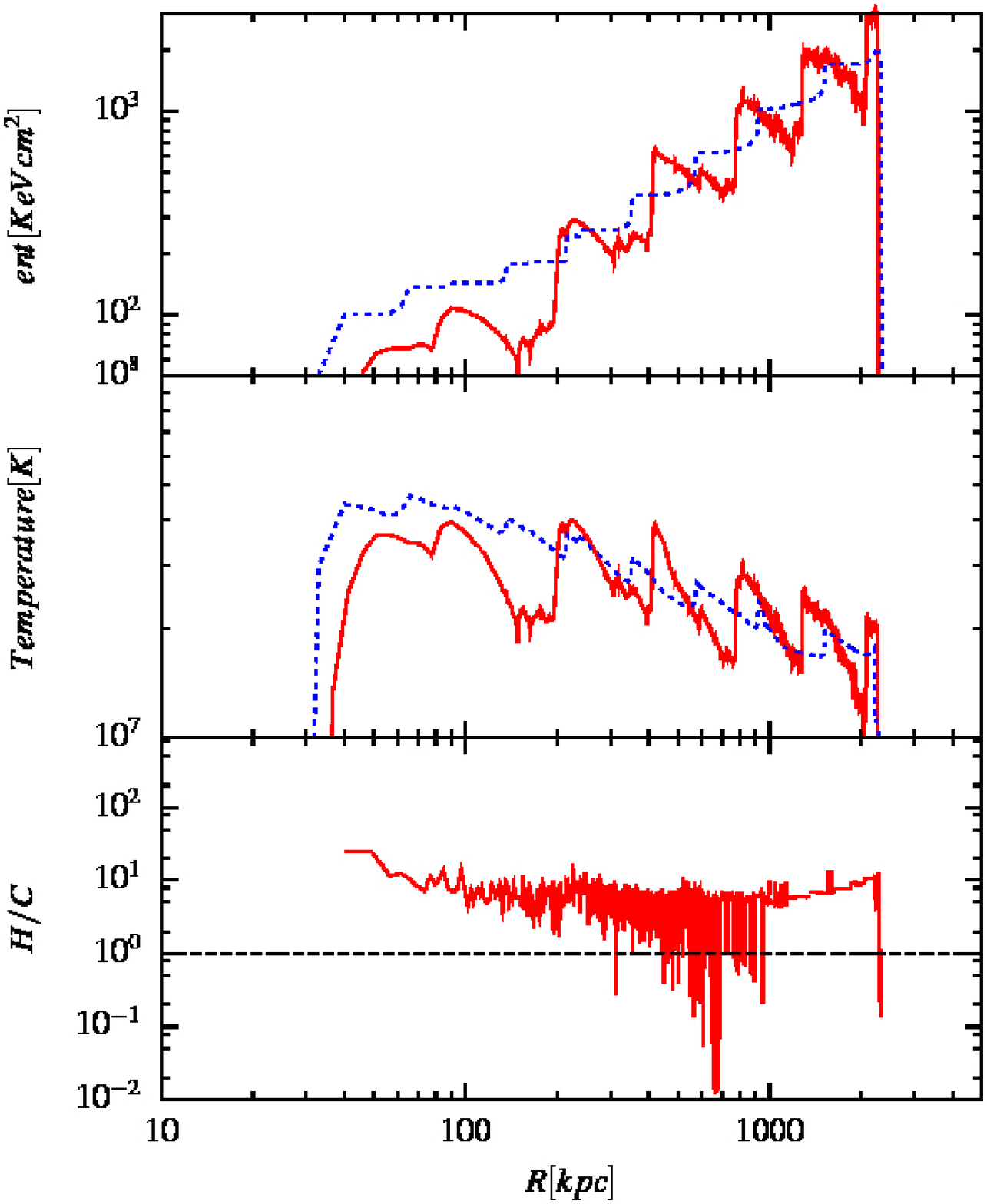}
\caption{Results of a 1D Hydrodynamic simulation of a $3\cdot
  10^{14}M_\odot$ halo with $f_c+f_b=0.1$ and
  $Z=0.3Z_\odot$. \emph{Left:} Flow lines of Lagrangian shells (grey
  thin lines, with temperature colormap. Top panel - no clumps, second
  panel - $f_c=0.01$ and $m_c=10^7M_\odot$, third panel - $f_c=0.01, m_c=10^7M_\odot$ with
  mixing length model for convection (``clump+convection''), bottom panel - mass
  evolution of the central galaxy
  (red; solid - ``no clumps'', green; long-dashed - $f_c=0.01$, blue;
  dashed - ``clump+convection'').
  \emph{Right:} Radial profiles at $z=0$ of the ``no clump'' (red; solid) and
  ``clump+convection'' (blue; dashed) simulations. top - entropy, middle -
  temperature. Bottom - H/C for the ``clump+convection'' case. 
  }
\label{fig:hydra}
\end{figure}
The evolution of galaxy and cluster halos from initial perturbation
has been studied with a 1D spherical Lagrangian hydrodynamic
simulation with 1D dark matter shells that evolve separately through
the baryonic shells, cooling and angular momentum
\citep{bd03,db06,bdn07}. The effect of clump heating was added to the
simulation as a sub-grid model of ``clump shells'' by assigning many
clumps to shells 
similar to dark matter shells and allowing them to interact with the
baryons via the drag equation and internal energy equation of the
baryons. Each clump shell contains $n$ clumps, and once the conditions for
disintegration of clumps are met the clump mass of that shell is
divided by $2$ and $n$ is multiplied by $2$. Once clumps are killed
(their mass drops below the conduction lower limit, $10^4M_\odot$) the
mass of the clump shell is added in situ to the baryonic shell overlapping the clump
shell, and the new temperature is calculated by mixing the hot halo
gas with the cold shell gas.
Since the gas is
cooling/heating unstable (see section \ref{sec:physics}), shells
heat and expend in an unstable manner. To deal with this physical
situation, gaseous shells are split into two when they become larger
than their surrounding shells (AMR), and physical convection is modeled by
mixing length theory \citep{spiegel63} was implemented.

Figure \ref{fig:hydra} shows results of three such simulations with
similar initial profile that leads to a $3\cdot 10^{14}M_\odot$ halo
at $z=0$. In the
first, no clumps are added, the CD galaxy grows to $3\cdot
10^{12}M_\odot$ and the star formation rates remains larger than
$100M_\odot/yr$ since $z=1$. By assuming that $10\%$ of the baryons
are accreted as clumps the size of the CD galaxy is reduced to
$1.5\cdot 10^{12}M_\odot$, star formation rate reduced to virtually
zero. The third simulation is similar to the second, except for mixing
length model is used to smooth over the local instabilities. In all
three simulations the virial mass and virial shock radius did not
change indicating that the heating does not effect the global
stability of the halo. The shape of the entropy profile and the
central values are consistent with those of cooling flow profiles
of\citet{donahue06}.
The hydrodynamic code and simulation results are described in detail
in \citetext{Birnboim \& Dekel 2009, in preparation}

\section{summary}
\label{sec:summary}
We have shown that a sufficient amount of energy is released in the
gravitational accretion to solve the overcooling problem of clusters
(larger than $M_{vir}\ge 10^{13}M_\odot$),
if we assume that some part of the baryons penetrates to the inner
parts rather that being stopped at the virial radius. The coupling
between the incoming accreted mass and the ambient gas is achieved by
assuming that the accreted baryons includes gaseous cold clump, that
penetrate through the virial shock and heat the gas via hydrodynamic
drag.
A parameter space survey indicates that for halos larger than
$10^{13}M_\odot$ and clump masses in the range $10^5\le m_c \le
10^8M_\odot$, clump heating can potentially solve the overcooling
problem.
The dynamic response of cluster halos to clump heating, cooling and
mass deposition in the inner parts in then examined.
For the set of values $M_{vir}=3 \cdot 10^{14}M_\odot$,
$m_c=10^7M_\odot$, $f_b+f_c=0.1$, $f_c=0.01$, the resulting entropy
and temperature profiles match typical observed cooling flow clusters
reasonably well. 
The physical processes discussed here are already included in 3D
hydrodynamic simulations but we note that to simulate the drag and the
formation of such clumps each $10^7M_\odot$ clump need to be well
resolved, which is typically not the case for $10^{15}M_\odot$ clusters.

\acknowledgements 
These proceedings summarize work carried out in collaboration with my
PhD advisor and current collaborator,
Avishai Dekel, to whom I wish to thank.

\end{document}